\documentclass[aps,pra, showpacs,twocolumn]{revtex4-1}
\usepackage{graphicx}
\usepackage{dcolumn}
\usepackage{bm}
\usepackage{color}
\usepackage{multirow}
\usepackage[mathscr]{euscript}
\usepackage{amsmath}

\begin{document}
\author{Weibin Li}
\author{Igor Lesanovsky}
\affiliation{School of Physics and Astronomy, University of Nottingham, Nottingham, NG7 2RD,
UK}
\title{Coherence in a cold atom photon switch}
\begin{abstract}
We study coherence in a cold atom single photon switch where the gate photon is stored in a Rydberg spinwave. With a combined field theoretical and quantum jump approach and by employing a simple model description we investigate systematically how the coherence of the Rydberg spinwave is affected by scattering of incoming photons. With large-scale numerical calculations we show how coherence becomes increasingly protected with growing interatomic interaction strength. For the strongly interacting limit we derive analytical expressions for the spinwave fidelity as a function of the optical depth and bandwidth of the incoming photon.
\end{abstract}
\date{\today}
\maketitle

\section{Introduction}
Cold gases of Rydberg atoms are currently receiving a growing attention in the communities of quantum optics~\cite{jonathan_d._pritchard_nonlinear_2012,peyronel_quantum_2012,gorshkov_dissipative_2013}, quantum information~\cite{saffman_quantum_2010}, and many-body physics~\cite{weimer_08,Stanojevic_09,weimer_rydberg_2010,pohl_crystal_10,lesanovsky_many-body_2011,Ji_11,garttner_12,petrosyan_13,Hu_13}. This is rooted in the fact that they offer strong and long-ranged interactions and at the same time grant long coherent lifetimes. Currently, considerable efforts are devoted to developing all-optical quantum information protocols~\cite{knill_scheme_2001,paredes-barato_all-optical_2014}  with the Rydberg-atom-mediated interaction between individual photons~\cite{petrosyan_quantum_2008,gorshkov_photon-photon_2011,he_two-photon_2014}. Fundamentally important optical devices that operate on the single photon level, such as phase shifters~\cite{firstenberg_attractive_2013}, switches~\cite{baur_single-photon_2014} and transistors~\cite{tiarks_single-photon_2014,gorniaczyk_single_2014}, have been demonstrated experimentally in Rydberg gases.

Single photon switchs might form a central building block of an all-optical quantum information processor~\cite{miller_2010,caulfield_2010,volz_2012}. The prime function of such switches is to control the transmission of an incoming photon through a single gate photon. One promising way to realize this is to store the gate photon in form of a gate (Rydberg) atom immersed in an atomic gas which is in a delocalized spinwave state~\cite{dudin_emergence_2012,wang_14,wang_15}. The gate atom then prevents transmission of incident photons through the gas, while ideally the coherence of the Rydberg spinwave state is preserved~\cite{duan_2002,porras_collective_2008,bariani_retrieval_2012,mirosh13}. The latter property would permit the subsequent coherent conversion of the Rydberg spinwave into a photon which would pave the way for gating the switch with superposition states that can also be subsequently retrieved. Currently, there is only a basic understanding of how the coherence of the Rydberg spinwave might be affected by the scattering of incoming photons and no systematic study of this important question exists.
\begin{figure}
\centering
\includegraphics*[width=0.97\columnwidth]{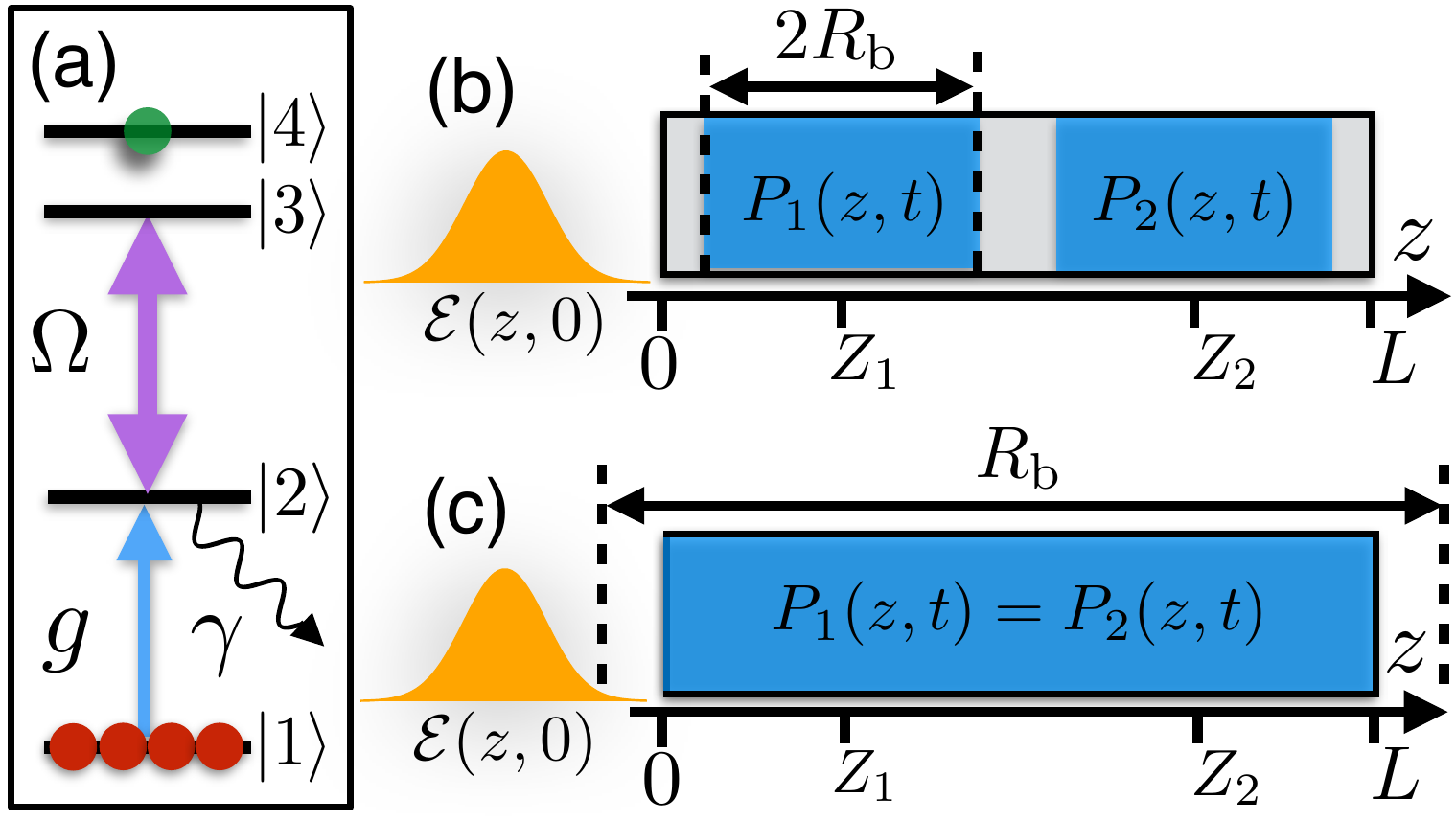}
\caption{(a) EIT level scheme. The groundstate $|1\rangle$, excited state $|2\rangle$ (decay rate $\gamma$) and Rydberg state $|3\rangle$ are resonantly coupled by a single photon field $\mathcal{E}(z,t)$ (with collective coupling strength $g$) and a classical field of Rabi frequency $\Omega$. Initially a gate photon is stored as a spinwave in the Rydberg state $|4\rangle$ (indicated by the green circle). (b,c) Polarization profiles $P_j(z,t)$ for a spinwave consisting of two possible gate atom positions $Z_j$ ($j=1,2$) and their dependence on the blockade radius $R_\mathrm{b}$ and the system length $L$. (b) For $L>R_{\text{b}}$ and $|Z_2-Z_1|>2R_{\text{b}}$ the polarization profiles associated with the two gate atom positions are distinguishable. (c) When $L\lesssim R_{\text{b}}$ the polarization profile is independent of the gate atom position which leads to enhanced coherence of the stored spinwave.}
\label{fig:illustration}
\end{figure}

In this work we address this outstanding issue within a simple model system. We study the propagation of a single photon under conditions of electromagnetically induced transparency (EIT) in a cold atomic gas in which a gate photon is stored as a Rydberg spinwave. An incident photon subsequently experiences a Rydberg mediated van der Waals (vdW) interaction with this stored gate atom which lifts the EIT condition and renders the atomic medium opaque. In this case the incident photon is scattered incoherently off the Rydberg spinwave. We study the photon propagation and explore the dependence of Rydberg spinwave coherence on the interaction strength (parameterized by the blockade radius $R_{\text{b}}$), the system length $L$ and bandwidth of the incident photon pulse. Our findings confirm that strong absorption, i.e. high gain, can be achieved already for large systems ($L>R_{\text{b}}$) while coherence of the spinwave is preserved only for sufficiently strong interactions, i.e. $L\lesssim R_{\text{b}}$. Intuitively, this can be understood by regarding the scattering of the incoming photon as a measurement of the position of the gate atom. When $L\lesssim R_{\text{b}}$ this measurement is not able to resolve the position of the excitation and hence coherence of the Rydberg spinwave is maintained. Our study goes beyond this simple consideration by taking into account propagation effects, a realistic interaction potential and a finite photon band width. The results can therefore be considered as upper bounds for the fidelity with which a Rydberg spinwave can be preserved and re-converted into a photon in an experimental realization of a coherent cold atom photon switch.

The paper is organized as follows. In section II, we introduce a one-dimensional model system to study the propagation dynamics of single source photons in the atomic gas prepared in a Rydberg spinwave state. In Sec. III, the model system is solved numerically with realistic parameters. We identify the working regime for a single photon switch where the source photon is scattered completely. In Sec. IV, we numerically study the fidelity between the initial spinwave state and the final state after the source photon is scattered. Our calculation shows that the coherence of the spinwave is preserved when $L \ge R_{\text{b}}$ while the final state becomes a mixed state when $L<R_{\text{b}}$. In Sec.~V, We provide analytical results for a coherent single photon switch ($L\ge R_{\text{b}}$). We reveal that the transmission and switch fidelity depend nontrvially on the optical depth and bandwidth of the source photon field. We summarize in Sec. VI. 
\section{the model system}
Our model system is a one-dimensional, homogeneous gas consisting of $N$ atoms, whose electronic levels are given in Fig.~\ref{fig:illustration}a. The photon field $\hat{\mathcal{E}}(z,t)$ and the EIT control laser (Rabi frequency $\Omega$) resonantly couple the groundstate $|1\rangle$ with the excited state $|2\rangle$ and $|2\rangle$ with the Rydberg state $|3\rangle$. Following Ref.~\cite{fleischhauer_quantum_2002}, we use polarization operators $\hat{P}(z,t)$ and $\hat{S}(z,t)$ to describe the slowly varying and continuum coherence of the atomic medium $|1\rangle\langle 2|$ and $|1\rangle\langle 3|$, respectively. All the operators $\hat{O}(z,t)=\{\hat{\mathcal{E}}(z,t)\,,\hat{P}(z,t)\,,\hat{S}(z,t)\}$ are bosons and satisfy the equal time commutation relation, $[\hat{O}(z,t),\hat{O}^{\dagger}(z',t)]=\delta(z-z')$. Initially, the atoms are prepared in a delocalized spinwave state with a single gate atom in the Rydberg state $|4\rangle$,
\begin{eqnarray*}
|\Psi_N(0)\rangle=\frac{1}{\sqrt{N}}\sum_{i=1}^Ne^{ikZ_i}|Z_i\rangle,
\end{eqnarray*}
where $k$ is the wavenumber of the spinwave and $|Z_i\rangle=|1_1\dots 4_i\dots 1_N\rangle$ abbreviates many-body basis with the gated atom located at position $Z_i$ and the rest in the groundstate. The Rydberg spinwave state  is created routinely in experiments~\cite{dudin_strongly_2012,li_entanglement_2013,baur_single-photon_2014,tiarks_single-photon_2014,gorniaczyk_single_2014}. When interacting with the incoming single photon, the general many-body state of this one-dimensional system is expanded as~\cite{peyronel_quantum_2012}
\begin{eqnarray}
|\Psi_N(t)\rangle&=&\left[\xi+\int dzE(z,t)\hat{\mathcal{E}}^{\dagger}(z,t)+\int dzP(z,t)\hat{P}^{\dagger}(z,t)\right.\nonumber\\
&+&\left.\int dzS(z,t)\hat{S}^{\dagger}(z,t)\right]|\Psi_N(0)\rangle,
\label{eq:state}
\end{eqnarray}  
where $\xi$ is probability amplitude of the initial spinwave state. In the weak field approximation, we will assume $\xi=1$ at any moment. We have defined $O(z,t)=\langle \hat{O}(z,t)\rangle$, i.e. the expectation value of the operator $\hat{O}(z,t)$. Specifically one finds that $E(z,t)$ is the probability amplitude in the one photon state, $P(z,t)$ and $S(z,t)$ are the amplitude of one atom in the $|2\rangle$ and $|3\rangle$ state, respectively. 

In order to develop a first intuition for the physics at work we first consider a spinwave that is delocalized merely over two atoms embedded in the atomic cloud (see Fig.~\ref{fig:illustration}b,c). We assume furthermore that the interaction between atoms in state $|3\rangle$ and the gate atom is infinite for distances smaller than the so-called blockade radius $R_\mathrm{b}$ and zero otherwise. Outside the blockade region, the photon propagates (along the $+z$ direction) as a dark-state polariton by virtue of EIT~\cite{fleischhauer_quantum_2002}. Inside the blockade region the medium behaves like an ensemble of two-level system. 
Here the incoming photon is building up a non-zero polarization $P(z,t)$, whose modulus square is the probability density distribution for finding an atom in the decaying state $|2\rangle$ according to Eq.~(\ref{eq:state})~\cite{interpretation}. Eventually, this leads to the loss of the incoming photon and makes the medium opaque. In order to understand how such photon scattering affects the coherence of the properties of the spinwave one needs to analyze the shape of the polarization profile. As shown in Fig.~\ref{fig:illustration}b this in general depends on the position of the gate atom when the system length is larger than the blockade radius $R_{\text{b}}<L$. Here, since $L>4R_{\text{b}}$ and $|Z_2-Z_1|>2R_{\text{b}}$, it is possible to distinguish the profiles  $P_j(z,t)$ which are associated with the two possible positions of the gate atom. Conversely, the polarization $P_j(z,t)$ becomes independent of the gate atom position when $L\lesssim R_{\text{b}}$ (see Fig.~\ref{fig:illustration}c). In this case --- as discussed in detail later --- the coherence of the spinwave will be preserved as one can not distinguish gate atoms from the scattered photon.

Let us now consider the actual photon propagation together with a realistic interaction potential. The dynamics of the system follows the master equation~\cite{fleischhauer_quantum_2002,plenio_quantum-jump_1998}
\begin{eqnarray}
\label{eq:masterequation}
\dot{\hat{\rho}}(t)=-i[\hat{H}_{\text{e}},\hat{\rho}(t)]+\gamma\int_0^L dz\hat{P}(z,t)\hat{\rho}(t)\hat{P}^{\dagger}(z,t),
\end{eqnarray}
where the first term on the right-hand side (RHS) is the evolution of $\hat{\rho}(t)$ under the effective Hamiltonian $\hat{H}_{\text{e}}=\hat{H}_\text{p}+\hat{H}_\text{ap}+\hat{H}_\text{a}$, and the spontaneous decay (with rate $\gamma$) from the state $|2\rangle$ is governed by the second term. In the effective Hamiltonian, the photon propagation in the medium is governed by the Hamiltonian
\begin{equation*}
\hat{H}_{\text{p}}= -c\int dz\hat{\mathcal{E}}^{\dagger}(z,t)\partial_z\hat{\mathcal{E}}(z,t),
\end{equation*}  
 with the vacuum light speed $c$. The atom-photon coupling is described by 
 \begin{eqnarray*}
\hat{H}_\text{ap}&=&-\int dz\left[\frac{i\gamma}{2} \hat{P}^{\dagger}(z,t)\hat{P}(z,t) + g\hat{\mathcal{E}}(z,t)\hat{P}^{\dagger}(z,t)\right.\\ &&+\left.\Omega\hat{S}^{\dagger}(z,t)\hat{P}(z,t) +\text{h.c.}\right],
 \end{eqnarray*}
  where $g=\sqrt{N}g_{\text{s}}$ with $g_{\text{s}}$ being the single atom-photon coupling strength. The vdW interaction between an atom in the state $|3\rangle$ and the gate atom at position $Z_i$ is
\begin{eqnarray*}
\hat{H}_\text{a}=\sum_i \int dz\hat{S}^{\dagger}(z,t)\hat{V}_i(z)\hat{S}(z,t).
\end{eqnarray*}
The interaction potential depends on the gate atom position,
\begin{equation*}
\hat{V}_i(z)= V_i(z) |Z_i\rangle\langle Z_i|,
\end{equation*}
where $V_i(z)  = C_6/(Z_i-z)^6$ gives the vdW interaction with $C_6$ being the dispersion coefficient.

For the case of a single incoming photon which we consider here the solution of the master equation (\ref{eq:masterequation}) is ~\cite{plenio_quantum-jump_1998}
\begin{equation}
\label{eq:densityevolution}
\hat{\rho}(t)=e^{-i\hat{H}_{\text{e}} t}\hat{\rho}_{\text{i}}e^{i\hat{H}_{\text{e}}^{\dagger} t} + \gamma\int\limits_0^L\int\limits_{0}^t dz dt'\hat{P}(z,t')\hat{\rho}_{\text{i}}\hat{P}^{\dagger}(z,t'),
\end{equation}
where $\hat{\rho}_{\text{i}}=|\Psi_N(0)\rangle\langle \Psi_N(0)|$ and $\hat{P}(z,t)=e^{i\hat{H}_{\text{e}}t} \hat{P}(z,0)e^{-i\hat{H}_{\text{e}}t}$. The first term on the RHS describes the unhindered photon propagation through the medium, while the second term accounts for the photon scattering, i.e. photon-loss from the medium.

\section{transmission of the source photon}
To calculate (\ref{eq:densityevolution}) we first treat the dynamics under the effective Hamiltonian in the Heisenberg picture. To this end we obtain the equation of motion for the expectation values $O(z,t)$ from the corresponding operator Heisenberg equation~\cite{gorshkov07}. Note, that due to the linearity of the equations we can moreover calculate the expectation value for each component $|Z_j\rangle$ of the Rydberg spinwave, i.e. each of the possible positions of the gate atom, separately. This yields the set of equations
\begin{subequations}
\label{eq:me}
\begin{align}
\partial_t\mathcal{E}_j(z,t)&=-c\partial_z\mathcal{E}_j(z,t)+ig P_j(z,t),\\
\partial_tP_j(z,t)&=-\frac{\gamma}{2}P_j(z,t)+ig\mathcal{E}_j(z,t)+i\Omega S_j(z,t),\\
\partial_tS_j(z,t)&=-iV_j(z)S_j(z,t)+i\Omega P_j(z,t).
\end{align}
\end{subequations}
where the index $j$ labels the respective spinwave component. Alternatively, these equations can be obtained from a Heisenberg-Langevin approach~\cite{gorshkov_photon-photon_2011}. We solve the coupled equations (\ref{eq:me}) through a Fourier transform yielding the formal solution for the polarization
\begin{equation}
\label{eq:pfft}
P_j(z,t)=\int_{-\infty}^{+\infty} d\omega \chi_j(z)\tilde{\mathcal{E}}_0(\omega)e^{-i\omega\mathcal{T}+i\frac{g}{c}\int_0^z dz' \chi_j(z')}.
\end{equation}
Here we have abbreviated $\mathcal{T}=t-z/c$ and introduced the electric susceptibility
\begin{eqnarray*}
 \chi_j(z)=g\frac{\omega-V_j(z)}{\Omega^2-[\omega-V_j(z)](\omega+i\gamma/2)}.
\end{eqnarray*}
From $\chi_j(z)$ one can actually extract the blockade radius as the critical distance at which the vdW interaction and the control laser are equally strong. This yields $R_{\text{b}}=|\gamma C_6/2\Omega^2|^{1/6}$~\cite{gorshkov_photon-photon_2011}.

The polarization (\ref{eq:pfft}) depends on the Fourier transform $\tilde{\mathcal{E}}_0(\omega)$ of the photon field at position $z=0$. To be specific we take the photon pulse to be a Gaussian at $t=0$ which is normalized in space, 
\begin{equation*}
\mathcal{E}(z,t=0)=\frac{1}{(\pi c^2\tau^2)^{1/4}}\exp\left[-\frac{(z-z_0)^2}{2c^2\tau^2}\right].
\end{equation*}
 Here $\tau$ is the temporal duration of the pulse and $z_0$ is the initial central position ($z_0\ll -c\tau$). The band width of the pulse is then given by $\Delta\omega=1/\tau$. Note, that it is generally not possible to evaluate the formal solution (\ref{eq:pfft}) analytically. Moreover, numerical calculations are challenging since the involved time and length scales span several orders of magnitude~\cite{note1}.

\begin{figure}
\centering
\includegraphics*[width=0.91\columnwidth]{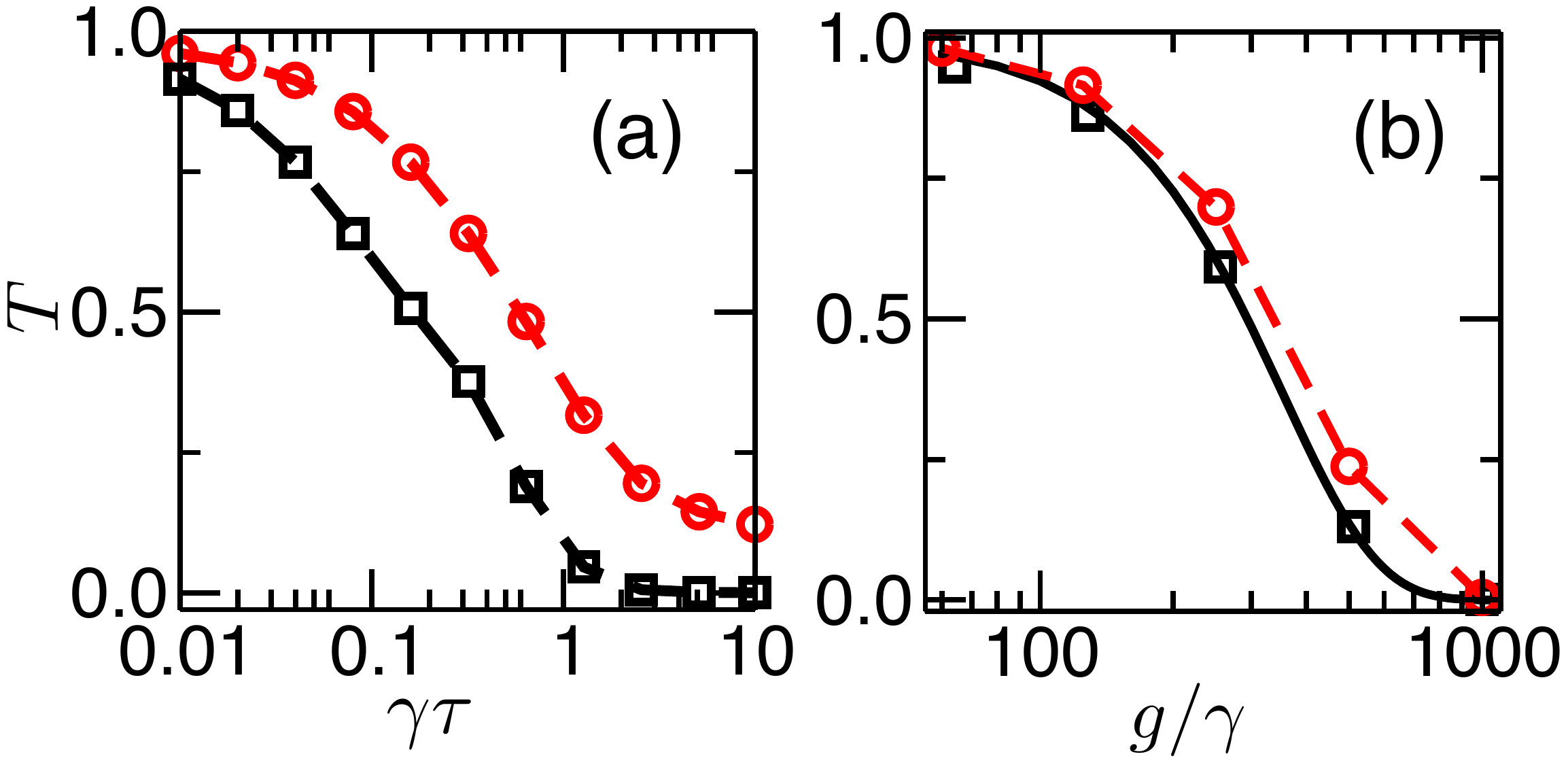}
\caption{(a) Photon transmission $T$ as a function of the pulse duration $\tau$ for $g=1000\gamma$ (squares) and  $g=500\gamma$ (circles). The medium becomes transparent when $\gamma\tau\ll 1$, i.e. the band width $\Delta \omega$ of the pulse is large. (b) Photon transmission as a function of the coupling constant $g$ for $R_{\text{b}}=L$ (squares) and $R_{\text{b}}=L/2$ (circles). The solid curve is the analytical result obtained from Eq.~(\ref{eq:transmission}). The dashed curve is plotted as a guide to the eye. Note, that $T\approx 0$ when $g= 1000\gamma$ for both $R_{\text{b}}=L$ and $R_{\text{b}}=L/2$. The data is calculated for rubidium atoms with the parameters, $L=20\,\mu$m, $\Omega=2\gamma$, and $\gamma\approx 2\pi\times 5.7$ MHz. The blockade radius can be changed through selecting different Rydberg states. For example, $R_{\text{b}}=L=20\,\mu$m when $|3\rangle=|127S\rangle$ and $|4\rangle=|130S\rangle$, where $C_6\approx 4.2\times 10^6\,\text{GHz}\,\mu \text{m}^6$.}
\label{fig:transmission}
\end{figure}

Let us now calculate the photon transmission as a function of the pulse duration $\tau$, which to our knowledge has not been examined previously. We define the transmission of the photon pulse as $T=\int_0^{\infty} dt |\mathcal{E}(L,t)|^2/\int_0^{\infty} dt |\mathcal{E}(0,t)|^2$. In Fig.~\ref{fig:transmission}a, we show $T$ as a function of the pulse width for two values of the atom-photon coupling strength $g$. For fixed pulse length $\tau$, we find that stronger couplings generally are accompanied by a lower transmission. Furthermore, we observe that the transmission increases with decreasing pulse duration $\tau$. This is due to the fact that the pulse contains increasingly more weight on frequency components, which are outside the absorption window of the medium. For the purpose of complete photon scattering, one thus has to utilize narrow frequency band pulses.

Next, we briefly discuss the dependence of the transmission $T$ on the strength of the atom-photon coupling $g$.  Fig.~\ref{fig:transmission}b shows data for two choices of the blockade radius, $R_{\text{b}}=L$ and $R_{\text{b}}=L/2$. As expected, $T$ decreases with increasing $g$. However, for the system parameters chosen here, there is virtually no dependence of $T$ on the value of the blockade radius when $g=1000\gamma$, where $T\approx 0$. These findings indicate that one reaches the strong scattering regime when $g\gg\gamma$ and $\Delta\omega\ll \gamma$. This is the working regime for the single photon switch where the medium becomes opaque for the incident photon.

\section{fidelity between the initial and final state}
Focusing on this regime, our next task is to investigate how the photon scattering influences the Rydberg spinwave. We quantify the difference between the initial Rydberg spinwave $\hat{\rho}_{\text{i}}$ and the final state $\hat{\rho}_{\text{f}}$ by the fidelity~\cite{uhlmann_transition_1976}
\begin{equation*}
F=\left[\text{Tr}|\sqrt{\hat{\rho}_{\text{i}}}\sqrt{\hat{\rho}_{\text{f}}}|\right]^2.
\end{equation*} 
As the initial spinwave is a pure state, this simplifies to $F=(1/N^2)\,\sum_{jk}A_{jk}$, where $A_{jk}=\gamma\int_0^L \int_0^{\infty} dz d\tau P_k^*(z,\tau)P_j(z,\tau)$. This shows that a high fidelity can be obtained only if the polarization profiles $P_j(z,\tau)$ for each spinwave component are essentially equal: Only when $A_{jk}\sim 1$ and thus $\sum_{jk}A_{jk} \sim N^2$ the fidelity is close to one. This is the formal version of the intuitive statement that we made earlier in conjunction with the discussion of Fig.~\ref{fig:illustration}b,c.

\begin{figure}
\centering
\includegraphics*[width=0.99\columnwidth]{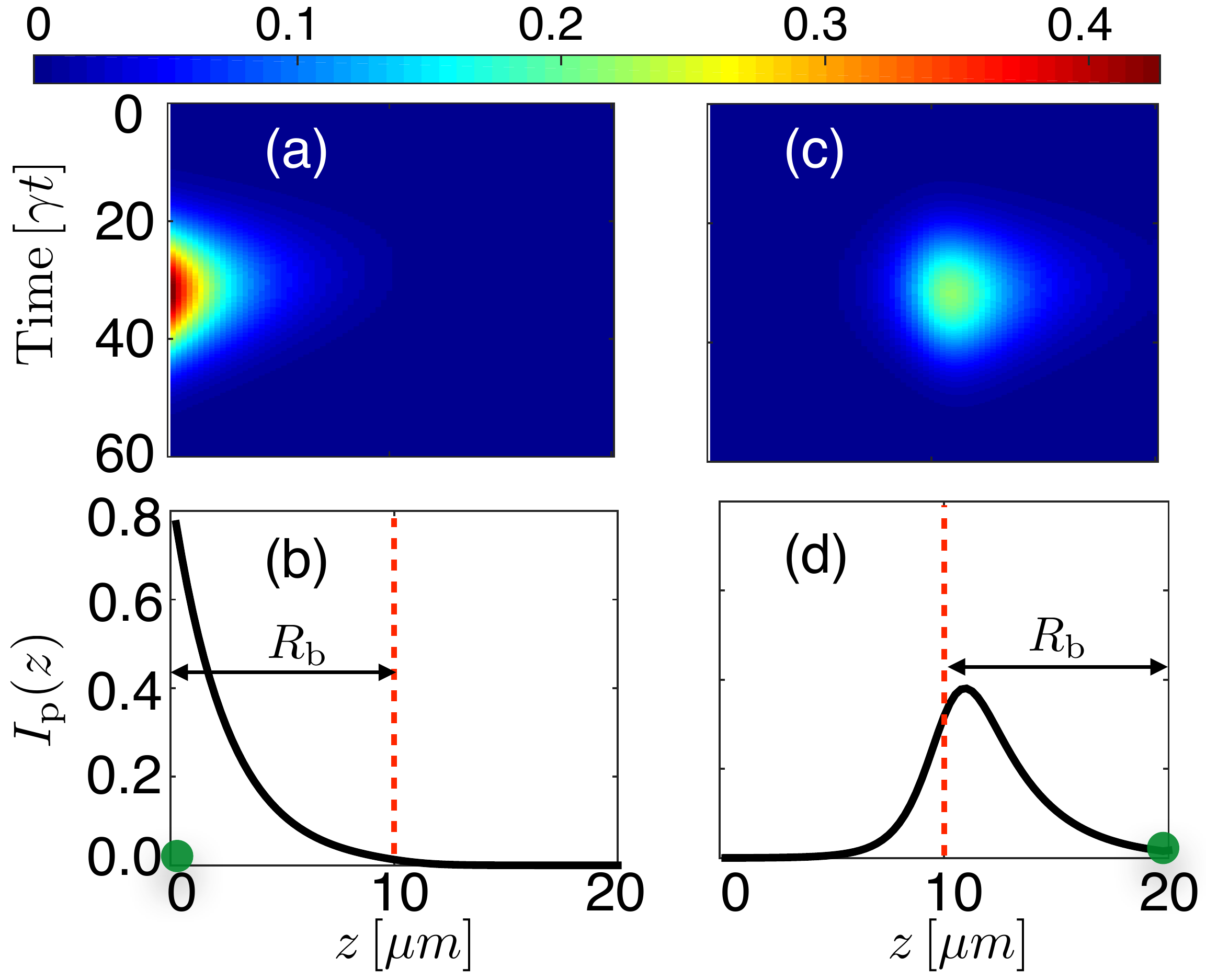}
\caption{(a,b) Squared modulus of the polarization $|P(z,t)|^2$ and (c,d) the time-integrated intensity $I_{\text{p}}(z)$ for $R_{\text{b}}=L/2$ and $g=1000\gamma$ for two different positions of the gate atom. The gate atom (green circle) is located at $Z_i=0$ in panels (a) and (b) and at $Z_i=L$ in panels (c) and (d). The dashed line marks the blockade radius with respect to the gate atom position.}
\label{fig:psi2}
\end{figure}

For completeness we provide a numerical example for which we choose $R_{\text{b}}=L/2$ and select only two components of the spinwave, where the gate atom is located at either $Z_i=0$ or $Z_i=L$. The resulting polarization profile $|P(z,t)|^2$ is shown in Fig.~\ref{fig:psi2}a,b. For $Z_i=0$, non-vanishing polarization is built up within the blockade region as long as the photon is inside the medium (Fig.~\ref{fig:psi2}a). Integrating over time we obtain the intensity $I_{\text{p}}(z)=\int_0^{\infty}d\tau |P(z,\tau)|^2 $ which clearly shows a decay to zero within a blockade distance $R_\mathrm{b}$ (see Fig.~\ref{fig:psi2}c). In contrast, for $Z_i=L$, appreciable polarization is built up also outside $R_\mathrm{b}$ and the profile is peaked at approximately $z=L-R_{\text{b}}$ (Fig.~\ref{fig:psi2}b,d). Clearly, both polarization profiles are strikingly different which in turn causes a loss of fidelity when the blockade radius is smaller than the system length. We verify this by numerically calculating the fidelity as a function of the blockade radius. The data is displayed in Fig.~\ref{fig:fidelity}a, together with the corresponding transmission $T$. As anticipated, the fidelity decreases significantly below unity when $R_{\text{b}}$ is decreased with respect to $L$. Note, that the transmission is close to zero throughout.

A fidelity smaller than unity directly indicates the formation of a mixed state after the photon scattering. The final state density matrix is
\begin{equation*}
\hat{\rho}_{\text{f}} =\frac{1}{N} \sum_{jk}A_{jk}e^{[ik(Z_j-Z_k)]} |Z_j\rangle\langle Z_k|.
\end{equation*}
The final state can only be pure when $|A_{jk}|=1$ and hence $F=1$. The formation of a mixed state is a consequence of the actual measurement of the gate atom position~\cite{gardiner_quantum_2004} which is performed by the photon scattering: When $R_{\text{b}}<L$ one in principle gains information on the position of the gate atom since the spatial uncertainty of its wave function is reduced from $L$ to the blockade region. The final state is then a mixture of all states compatible with this additional information.

\section{analytical results for coherent photon switchs}
\begin{figure}
\centering
\includegraphics*[width=0.91\columnwidth]{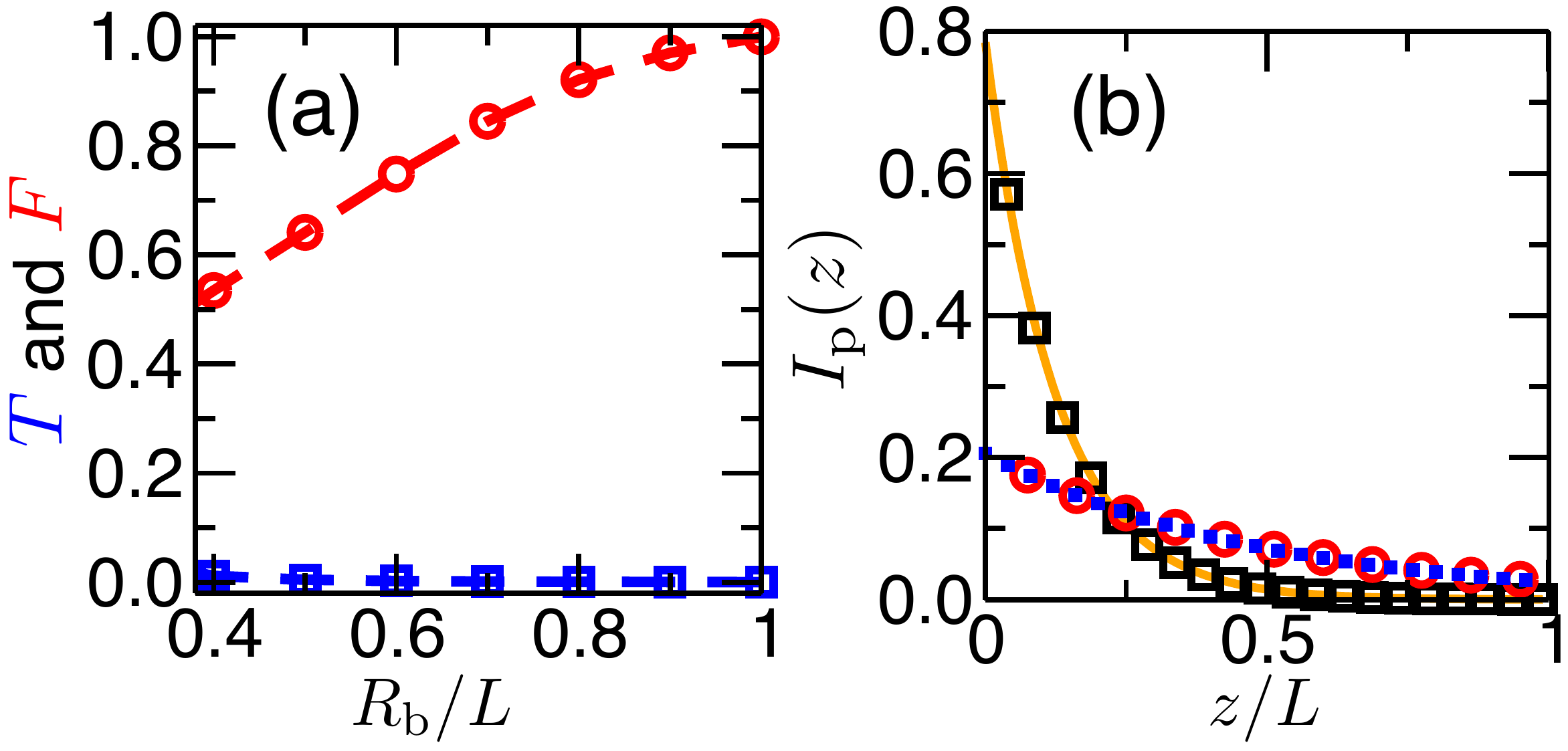}
\caption{(a) Fidelity (circles) and transmission (squares) as a function of the blockade radius $R_{\text{b}}$ with $g=1000\gamma$. The transmission is already negligible when $R_{\text{b}}/L>0.4$, but the fidelity approaches unity only when $R_{\text{b}}\sim L$. (b) Intensity $I_{\text{p}}(z)$ in the strong blockade regime ($R_{\text{b}}=L$) for $g=1000\gamma$ (square) and $g=512\gamma$ (circle). The squares and circles are numerical data. The solid and dotted curves are the analytical results obtained from Eq.~(\ref{eq:polarization_response}).}
\label{fig:fidelity}
\end{figure}

In the remainder of the paper we will focus on the case of a coherent photon switch, i.e. $R_{\text{b}}\geq L$. Here the expression for the susceptibility of the medium simplifies to that of an ensemble of two-level atoms, $\chi_j(z)\approx -g/(\omega+i\gamma/2)$ which permits the derivation of analytical results. For a narrow band width pulse we can derive explicit solutions to Eq.~(\ref{eq:me}) that have no dependence on the position of the gate atom~\cite{supp}. For example, the polarization $P(z,t)$ is given by
\begin{eqnarray}
\label{eq:polarization_response}
P(z,t) &\approx& ig\sqrt{\frac{\sqrt{\pi}\tau}{2c}}\exp\left[\frac{c\gamma^2\tau^2-4\gamma(c\mathcal{T}+ z_0)}{8c}-\frac{4g^2z}{c\gamma}\right]\nonumber\\
&&\times \text{Ec}\left[\frac{c\gamma^3\tau^2-2\gamma^2(c\mathcal{T}+z_0)-8g^2z}{2\sqrt{2}c\tau\gamma^2} \right],
\end{eqnarray}
where $\text{Ec}(x)$ is the complementary error function. The corresponding time-integrated profile $I_{\text{p}}(z)$ agrees perfectly with the numerical result from Eq.~(\ref{eq:me}) (see Fig.~\ref{fig:fidelity}b). The transmission $T$ is given by
\begin{equation}
\label{eq:transmission}
T\approx e^{-\alpha} \left[\text{Ec}\left(\frac{z_0}{c\tau}\right)\right]^{-1}\left[1+\text{Er}\left(\frac{L-z_0}{c\tau}-\frac{\alpha}{\gamma\tau}\right)\right],
\end{equation}
where $\text{Er}(x)$ is the error function and $\alpha=4g^2L/c\gamma$ is the optical depth of a resonant two-level medium. The excellent agreement between the analytical and numerical calculation is shown in Fig.~\ref{fig:transmission}b. Neglecting the finite band width of the photon pulse, i.e. when all the frequency components are in the absorption window, Eq.~(\ref{eq:transmission}) reduces to the well-known form $T\approx e^{-\alpha}$~\cite{gorshkov_photon-photon_2011}.

Finally, the fidelity can be expressed as a function of the optical depth and pulse band width
\begin{equation}
\label{eq:fidelity_narrowband}
F\approx (1-e^{-\alpha}) \left[1-\frac{2}{\gamma^2\tau^2}+\frac{12}{\gamma^4\tau^4}\right].
\end{equation}
This shows that indeed a small band width is a requirement for reaching a large fidelity. For example, the transmission is negligible ($T\approx 8\times 10^{-4}$) when $\gamma\tau=5$ and $g=1000\gamma$ according to the data in Fig.~\ref{fig:transmission}a. However, the fidelity is below unity ($F=0.94$) due to non-negligible contributions from the terms accounting for the finite band width. In the limit of very long pulses one finds $F \approx 1-T$ and thus the fidelity is solely determined by the transmission.
\section{summary}

In summary, we have studied the coherence of a Rydberg spinwave in the operation of  a signle photon switch. The current study is limited to a single gate atom and an incoming single-photon pulse, which permits the description of multi-photon scattering, however, only if the photons enter the switch sequentially. Addressing this limitation and extending the discussion to correlated and entangled photon pulses that fall in the operation regime of single photon transistors will be subject to future studies.

\begin{acknowledgments}
\textit{Acknowledgements.---} We acknowledge helpful discussions with D. Viscor, B. Olmos and M. Marcuzzi. The research leading to these results has received funding from the European Research Council under the European Union's Seventh Framework Programme (FP/2007-2013) / ERC Grant Agreement No. 335266 (ESCQUMA), the EU-FET grants No. 295293 (QuILMI) and No. 512862 (HAIRS), as well as the H2020-FETPROACT-2014 Grant No. 640378 (RYSQ). WL is supported through the Nottingham Research Fellowship by the University of Nottingham.
\end{acknowledgments}

\appendix
\section{Details of the analytical calculation}
 Here we will show how to obtain the analytical solution to the coupled equations (3) in the main text in the strong blockade regime ($R_b\ge L$) and for narrow band pulses. 

In the frequency domain, the solution to equations (3) is given by,
\begin{eqnarray}
\label{eq:e}
\tilde{\mathcal{E}}(z,\omega)&=&\exp\left[\frac{i\omega}{c}z+\frac{igz}{c} \chi \right]\tilde{\mathcal{E}}_0(\omega),\\
\label{eq:p}
\tilde{P}(z,\omega)&=&\chi\tilde{\mathcal{E}}(z,\omega),\\
\label{eq:s}
\tilde{S}(z,\omega)&=&-\frac{\Omega}{\omega-V(z)}\chi\tilde{\mathcal{E}}(z,\omega).
\end{eqnarray}
Due to the strong blockade condition, we have removed the dependence of $\tilde{\mathcal{E}}(z,\omega)$, $\tilde{P}(z,\omega)$, and $\tilde{S}(z,\omega)$ on the gate atom index $j$ and replaced the susceptibility by the one corresponding to two-level atoms, $\chi=-g/(\omega+i\gamma/2)$. Moreover, we set $\tilde{S}(z,\omega)\approx 0$, which is a good approximation as $\Omega/[\omega-V(z)]\approx \Omega/V(z)\approx 0$ in Eq.~(\ref{eq:s}). Our aim is to obtain analytical expressions of $\mathcal{E}(z,t)$ and $P(z,t)$.

Applying the inverse Fourier transform on the both sides of Eqns.~(\ref{eq:e}) and (\ref{eq:p}), we obtain the formal solution for $\mathcal{E}(z,t)$ and $P(z,t)$ in time domain ,
\begin{eqnarray}
\label{eq:eifft}
\mathcal{E}(z,t)&=&\int_{-\infty}^{\infty}d\omega\tilde{\mathcal{E}}_0(\omega)e^{-i\omega\mathcal{T}+i\frac{gz}{c}\chi},\\
\label{eq:psi2ifft}
P(z,t)&=&\int_{-\infty}^{\infty} d\omega \chi\tilde{\mathcal{E}}_0(\omega)e^{-i\omega\mathcal{T}+i\frac{gz}{c} \chi}.
\end{eqnarray}
The integration is in general difficult to carry out analytically due to the complicated form of the susceptibility. We overcome this difficulty by expanding the susceptibility in powers of $\omega$,
\begin{equation}
\label{eq:s_expansion}
\chi=\frac{2g}{\gamma}\left[i-\frac{2\omega}{\gamma} -\frac{4i\omega^2}{\gamma^2}+\cdots\right].
\end{equation}
\begin{widetext}
First let us calculate the approximate solution for $\mathcal{E}(z,t)$. To carry out analytical calculations and at the same time take into account contributions due to the finite band width, we will keep terms up to the second order of $\omega$ in Eq.~(\ref{eq:s_expansion}). This yields the solution for $\mathcal{E}(z,t)$
\begin{eqnarray}
\label{eq:efield}
\mathcal{E}(z,t)&=&\frac{1}{\sqrt{\sqrt{\pi}c\tau}\xi(z)} \exp\left[-\frac{2g^2z}{c\gamma}-\frac{1}{2\xi^2(z)\tau^2}\left(t-\frac{\gamma^2-4g^2}{\gamma^2}\frac{z}{c}+\frac{z_0}{c}\right)^2\right],
\end{eqnarray}
with $\xi(z) = \sqrt{1-4\alpha z/L\gamma^2\tau^2}$. For the current problem, we always have $\xi(z)>0$ as the photon travelling time through the medium is the shortest time scale. For example, $L/c\approx 3\times 10^{-14}$ second for $L=10\,\mu$m.

With the solution for $\mathcal{E}(z,t)$, we can calculate the transmission $T=\int_0^{\infty} dt |\mathcal{E}(L,t)|^2/\int_0^{\infty} dt |\mathcal{E}(0,t)|^2$. We need to carry out the respective two integrals over time at $z=0$ and $z=L$. This can be done analytically,
\begin{equation}
	\int_0^{\infty} dt |\mathcal{E}(0,t)|^2=\frac{1-\text{Er}(\frac{z_0}{c\tau})}{2c},
\end{equation}
and 
\begin{equation}
	\int_0^{\infty} dt |\mathcal{E}(L,t)|^2=\frac{e^{-\alpha}}{2c\xi(L)}\left[1+\text{Er}\left(\frac{L-z_0}{c\tau\xi(L)}-\frac{\alpha}{\gamma\tau\xi(L)}\right)\right]\approx \frac{e^{-\alpha}}{2c\xi(L)}\left[1+\text{Er}\left(-\frac{z_0}{c\tau\xi(L)}-\frac{\alpha}{\gamma\tau\xi(L)}\right)\right].
\end{equation}
This leads to the analytical form of the transmission (7) in the main text.

With the analytical solution for $\mathcal{E}(z,t)$ at hand, there are two ways to calculate $P(z,t)$. We can directly calculate $P(z,t)$ from Eq.~(3b) by inserting the solution~(\ref{eq:efield}) and $S(z,t)=0$. This yields the linear response of the medium to the photon electric field,
\begin{equation}
	P(z,t)=ig\int_{-\infty}^td\tau e^{-\gamma (t-\tau)/2}\mathcal{E}(z,\tau). 
\end{equation}
The integration over time can be carried out analytically, which gives
\begin{eqnarray}
\label{eq:p_response}
P(z,t) &\approx& ig\sqrt{\frac{\sqrt{\pi}\tau}{2c}}\exp\left[\frac{c\gamma^2\tau^2-4\gamma(c\mathcal{T}+ z_0)}{8c}-\frac{6g^2z}{c\gamma}\right] \left[1+\text{Er}\left(\frac{1-3\xi^2(z)}{4\sqrt{2}\xi(z)}\gamma\tau +\frac{c\mathcal{T}+z_0}{\sqrt{2}\xi(z)c\tau} \right)\right].
\end{eqnarray}

However it is difficult to calculate the fidelity from Eq.~(\ref{eq:p_response}) due to the presence of the error function. We thus calculate $P(z,t)$ alternatively using the Fourier transform method. We note that the susceptibility $\chi$ appears at two places in Eq.~(\ref{eq:psi2ifft}): one in front of $\tilde{\mathcal{E}}(\omega)$ and another one in the exponential function.  In order to obtain an analytical result, we will expand the former susceptibility up to the second order of $\omega$ while the latter up to the linear order. After performing the inverse Fourier transform, we obtain the expression for $P(z,t)$,
\begin{eqnarray}
\label{eq:psi2_fft2nd}
P(z,t)&=&\frac{i g \mathcal{E}(z,t)}{32c^2\tau^4\gamma^9}\left\{c^2 \gamma ^6 \tau ^4+8 c z \gamma ^3 \tau ^2 g^2 -4 c^2 \gamma ^3 \tau ^2+ 2 c \gamma ^5 \tau ^2   (c t-z+z_0)+4 \left[\gamma ^2 (c t-z+z_0)+4 g^2 z\right]^2\right\}.
\end{eqnarray}
\end{widetext}

Using Eq.~(\ref{eq:psi2_fft2nd}) the fidelity can be calculated analytically, 
\begin{equation}
F_a\approx (1-e^{-\alpha}) \left[1-\frac{2}{\gamma^2\tau^2}+\frac{12}{\gamma^4\tau^4}\right].
\end{equation}


\begin{thebibliography}{43}
\bibitem [{not()}]{note1}%
\BibitemOpen
\href@noop {} {}\bibinfo {note} {Typically the pulse length $c \tau$ is
	thousands of meters, while the length of the atomic cloud is merely tens of
	micrometers. Due to the vacuum light speed $c$, the travelling time in the
	blockade region is hardly longer than 1 picosecond. This is far shorter than
	the time scale of the laser-atom interaction, which is in the order of
	microsecond. We use a Lax-Wendroff algorithm to solve the coupled
	differential equations. In order to have a good resolution in both time and
	space, we choose the spatial grid $dz=0.01 L$ and restrict the time step with
	the condition $dt<dz/c$ to ensure the numerical stability.}\BibitemShut
{Stop}%
\end{thebibliography}
\end{document}